\documentclass[showpacs,twocolumn,floatfix,amsmath,amssymb,superscriptaddress,prb,nopacs]{revtex4-1}
\usepackage{graphicx}
\usepackage{float}
\usepackage{dcolumn}
\usepackage{bm}
\usepackage{amssymb}
\usepackage{amsmath}
\usepackage{hyperref} 
\hypersetup{bookmarks=true,unicode=true,colorlinks=true, urlcolor=blue,linkcolor=blue,citecolor=blue,filecolor=blue}

\newcommand{\bt}{\textbf}
\usepackage{physics}
\usepackage{comment}
\graphicspath{{figures/}}

\begin{document}

\title{Spin-orbital polarization of Majorana edge states in oxides nanowires}

\author{J. Settino}
\affiliation{CNR-SPIN c/o Universit\'a degli Studi di Salerno, I-84084 Fisciano (Sa), Italy}
%
\author{F. Forte}
\affiliation{CNR-SPIN c/o Universit\'a degli Studi di Salerno, I-84084 Fisciano (Sa), Italy}
\affiliation{Dipartimento di Fisica ``E. R. Caianiello", Universit\'a degli Studi di Salerno, I-84084 Fisciano (Sa), Italy}
\author{C. A. Perroni}
\affiliation{CNR-SPIN c/o
Universit\'a degli Studi di Napoli Federico II,
\\ Complesso Universitario Monte S. Angelo, Via Cintia, I-80126 Napoli, Italy}
\affiliation{Physics Department "Ettore Pancini",
Universit\'a degli Studi di Napoli Federico II,
\\ Complesso Universitario Monte S. Angelo, Via Cintia, I-80126 Napoli, Italy}

\author{V. Cataudella}
\affiliation{CNR-SPIN c/o
Universit\'a degli Studi di Napoli Federico II,
\\ Complesso Universitario Monte S. Angelo, Via Cintia, I-80126 Napoli, Italy}
\affiliation{Physics Department "Ettore Pancini",
Universit\'a degli Studi di Napoli Federico II,
\\ Complesso Universitario Monte S. Angelo, Via Cintia, I-80126 Napoli, Italy}
\author{M. Cuoco}
\affiliation{CNR-SPIN c/o Universit\'a degli Studi di Salerno, I-84084 Fisciano (Sa), Italy}
\affiliation{Dipartimento di Fisica ``E. R. Caianiello", Universit\'a degli Studi di Salerno, I-84084 Fisciano (Sa), Italy}
\author{R. Citro}
\affiliation{CNR-SPIN c/o Universit\'a degli Studi di Salerno, I-84084 Fisciano (Sa), Italy}
\affiliation{Dipartimento di Fisica ``E. R. Caianiello", Universit\'a degli Studi di Salerno, I-84084 Fisciano (Sa), Italy}

\begin{abstract}
We investigate a paradigmatic case of topological superconductivity in a one-dimensional nanowire with $d-$orbitals and a strong interplay of spin-orbital degrees of freedom due to the competition of orbital Rashba interaction, atomic spin-orbit coupling, and structural distortions. We demonstrate that the resulting electronic structure exhibits an orbital dependent magnetic anisotropy which affects the topological phase diagram and the character of the Majorana bound states (MBSs). 
The inspection of the electronic component of the MBSs reveals that the spin-orbital polarization generally occurs along the direction of the applied Zeeeman magnetic field, and transverse to the magnetic and orbital Rashba fields.
The competition of symmetric and antisymmetric spin-orbit coupling remarkably leads to a misalignment of the spin and orbital moments transverse to the orbital Rashba fields, whose manifestation is essentially orbital dependent. 
The behavior of the spin-orbital polarization along the applied Zeeman field reflects the presence of multiple Fermi points with inequivalent orbital character in the normal state.  Additionally, the response to variation of the electronic parameters related with the degree of spin-orbital entanglement leads to distinctive evolution of the spin-orbital polarization of the MBSs. 
These findings unveil novel paths to single-out hallmarks relevant for the experimental detection of MBSs.
\end{abstract}

\maketitle

\section{Introduction}

One-dimensional topological superconductivity of intrinsic \cite{Sigrist,ReadGreen,Ivanov,KitaevUnpaired,VolovikBook,Maeno2012,SatoAndo} or artificial~\cite{FuKane,HasanKane,QiZhang,Alicea,CarloRev,Leijnse,KotetesClassi,Franz,Aguado,
LutchynNatRevMat} p-wave superconductors can harbor the so-called Majorana bound states (MBSs) which are pinned to zero-energy and are charge neutral.
Recent experimental observations~\cite{Yazdani1,Ruby,Meyer,Jinfeng,Yazdani2,Wiesendanger,Wray2010,Zhang2018} have brought evidences for the presence of 
MBSs in artificial topological superconductors, which are based on nanosized chains with magnetic atoms deposited on top of a superconducting substrate. In these experiments, while the MBSs occur in an effective spinless regime, their observed fingerprints arise from the nontrivial spin structure of the corresponding MBS wavefunction. There, the spin polarization of the MBS configuration can be accessed by means of scanning tunneling microscopy (STM) through a measurement of spin-selective conductance \cite{He,KotetesSpin,Maiellaro}. This physical scenario applies also to semiconducting nanowires proximity-coupled with an $s$-wave spin-singlet superconductor \cite{Wimmer2010,Potter2010,Lutchyn2011,Lim2012,Stanescu} and networks \cite{Alicea2011} where the electronic spin orientation and spatially resolved texture of the MBS can exhibit fingerprints that depend on the relative strength of Rashba and Dresselhaus interactions \cite{Sticlet2012}.
On a general ground, due to symmetry constraints a Kramers pair of MBSs is marked by an Ising spin, i.e., the spin density is nonvanishing only along a specific direction \cite{IsingSpin}. On the contrary, the spin density for the case of MBSs protected by a sublattice (chiral) symmetry is identically zero \cite{Sticlet2012,Mercaldo2019}. However, the electron projection of the spin-density is generally nonvanishing even for effective spinless topological superconductors, where the spin polarization is locked along a given orientation and this can be probed by STM or charge transport measurements. Along this line, a radically different situation can occur in intrinsic quasi one-dimensional topological superconductors, where the electron spin is an active degree of freedom in setting out the topological behavior \cite{Mercaldo2016,Mercaldo2017,Mercaldo2018}, and chiral protected multiple MBSs at the edges can manifest both an Ising type behavior and a spin texture with characteristic spatial patterns and orientations \cite{Mercaldo2019}.

\begin{figure*}[t!]
\centering
\includegraphics[height=11cm,width=0.85\textwidth]{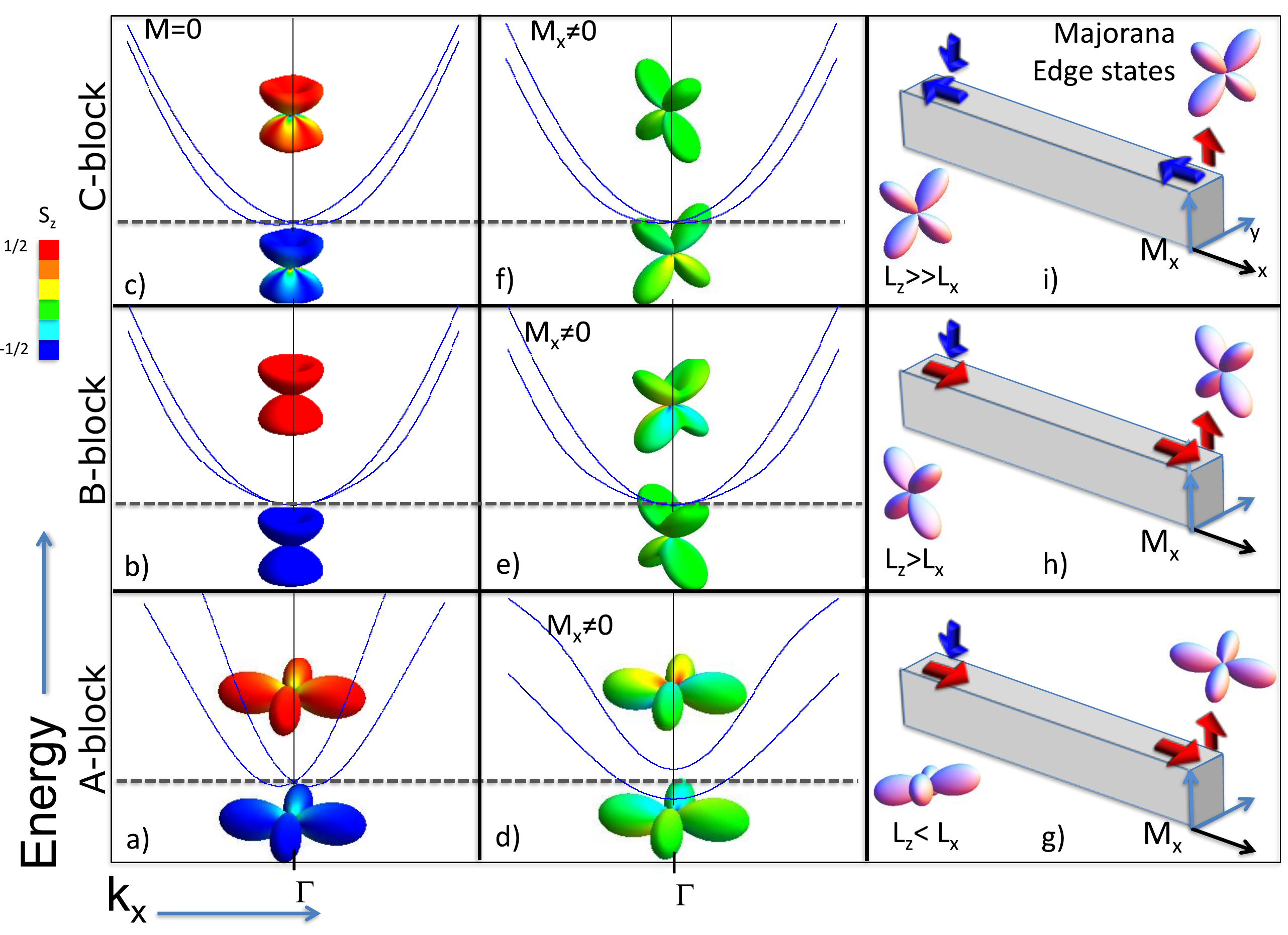}
\protect\caption{Schematic view nearby the $\Gamma$ point in the Brillouin zone of the bands arising from the considered atomic ($d_{xy},d_{xz},d_{yz}$)  configurations.  
The tetragonal crystal field potential splits the $d_{xy}$ with respect to ($d_{xz},d_{yz}$) lowering the energy of the $d_{xy}$ state (a). Then, the spin-orbit coupling leads to a configuration with nontrivial combination of spin ($s$)  and orbital ($l$) angular momentum as schematically illustrated for the $\Gamma$ point Kramers states in (a),(b) and (c). We notice that the states in (b) have only $z$-component of $l$, while the configurations in (a) and (c) have dominant $l_x$ and $l_z$ components, respectively. There, the degree of mixing can be qualitatively extracted from the inspection of the angular distribution. The application of a magnetic field splits the Kramers states at the $\Gamma$ point but due to the spin-orbit coupling the splitting amplitude depends on the orbital character. For a magnetic field applied along the nanowire direction ($x$) the lowest energy band has a larger splitting compared with the other bands due to the dominant spin-orbital polarization along $x$. Once the topological state is achieved for a given electron filling (dotted line indicates the chemical potential) the Majorana bound states (MBSs) occur at the edges of the nanowire with a characteristic spin-orbital content as sketched in (g), (h), and (i). We notice that the spin and orbital polarizations of the MBS lie in the $xz$ plane coplanar to the direction of the applied magnetic field and perpendicular to the orbital Rashba field direction (i.e. $y$). The behavior of the components are orbital dependent.}
\label{Fig:1}
\end{figure*}

Moving to a broader physical scenario, one can ask whether for superconducting materials having an electronic structure with nontrivial spin-orbital entanglement, the electron spin and orbital moment are active degrees of freedom in the MBSs and can leave a unique imprint on spin-resolved and, potentially, on orbitally-resolved local spectroscopic probes.   
Such appealing perspective can naturally occur when considering superconducting materials with atomic multi-orbital degrees of freedom. These microscopic elements are commonly encountered in oxides materials, where $d$-bands can lead to fascinating spin-orbital correlated phenomena which, apart from fundamental physical challenges, can also lead to tantalizing solutions for emergent technologies \cite{roadmap}.  
In this framework, a prototype superconductivity with electronic components having intrinsically coupled spin-orbital degrees of freedom is provided by the two dimensional electron gas (2DEG) forming at the interface of oxide band insulators. Oxide 2DEGs, indeed, are characterized  by the simultaneous presence of strong spin-orbit coupling \cite{Caviglia2010} and superconductivity \cite{Reyren2007}, both widely tunable by electric field effect \cite{Caviglia2008,Caviglia2010}, while 2D magnetism, coexisting with superconductivity, \cite{Hwang2012,Pai2018a} can be induced by opportune atomic engineering of the heterostructures \cite{Stornaiuolo2016}. Hence, the combination of magnetism, superconductivity and inversion asymmetry provides a quite unique platform for the realization of multi-orbital topological superconducting phases. Recently, quasi-2D electron gases formed at the interface between LaAlO$_3$ and SrTiO$_3$ (LAO/STO) \cite{Ohtomo2004} have been theoretically proposed as possible candidates for the realization of topological superconducting phases in two-dimensional \cite{Scheurer2015,Mohanta2014,Loder2015,Fukaya2018} and various topological scenarios have been explored in effective quasi one-dimensional models \cite{Fidkowski2011,Fidkowski2013,Mazziotti2018,Perroni2019,Joshua2012,Pai2018,Shalom}. 

In this work we aim to assess the spin-orbital character of MBSs occurring in a topological superconducting phase that is induced by an applied Zeeman magnetic field for a one-dimensional nanowire with $d-$orbitals ($d_{xy},d_{xz},d_{yz}$) and strong interplay of spin-orbital degrees of freedom. As the $d$-orbitals have a nontrivial angular momentum and an anisotropic spatial distribution, the nature of the electronic states is significantly sensitive to spin-orbit coupling and system's dimensionality. Here, we focus on a typical electronic situation in low-dimensional materials where the interplay of spin-orbit coupling and tetragonal distortions breaks the spin and orbital rotational invariance resulting into a characteristic atomic spin-orbital distribution (Figs. \ref{Fig:1}(a)-(c)). Additionally, inversion asymmetry yields orbital Rashba-type interaction that together with the spin-orbit coupling sets out a non-trivial momentum dependent spin-orbital splitting. In this framework, apart from the nonstandard spin-orbital texture naturally occurring at the Fermi level \cite{Fukaya2019}, the response to a Zeeman magnetic field is highly anisotropic and orbital dependent (Figs. \ref{Fig:1}(d)-(f)). Thus, once a topological superconducting phase is obtained, the emergent MBSs may exhibit unique fingerprints of the underlying spin-orbital electronic substrate from which they arise. This is indeed the key outcome of the paper and we find distinct characteristics of the spin-orbital content of the MBSs that we summarize in Figs. \ref{Fig:1}(g)-(i). The inspection of the electronc component of the MBSs reveals that the spin-orbital polarization has always a planar orientation. Moreover, the components along the direction of the applied Zeeeman magnetic field and orthogonal to the magnetic-and-orbital Rashba fields are strongly sensitive to a variation of the spin-orbit strength. The emerging trend is to have a tunable orientation which is dependent on the orbital character of the bands where the topological pairing sets in.
We also find that the competition of symmetric and antisymmetric spin-orbit coupling remarkably leads to a misalignment of the spin and orbital moment orientations for the MBSs whose manifestation is inequivalent for the $d_{xy}$ compared to the $d_{xz},d_{yz}$ based bands. Additionally, even in a regime where the spin-momentum locking substantially deviates from that due to the spin Rashba coupling, we find that the spin-orbital polarization has a planar orientation. We also investigate the behavior of the electron spin-orbital polarization along the applied Zeeman field across the topological phase transition and show that it reflects the presence of multiple Fermi points with inequivalent orbital character in the normal state. These findings unveil a rich scenario concerning the spin-orbital content of the MBSs and nonstandard paths to single-out hallmarks which may be relevant for the experimental detection of MBSs.

The paper is organized as follows. In Section II, the model hamiltonian for oxides nanowires is presented.
In Section \ref{sec:majorana-polarization}, we introduce the orbital dependent Majorana fermion polarization and present the topological phase diagram resulting from the application of a Zeeman magnetic field. 
In Section IV we provide the key fingerprints of the orientation and spatial profile of the electron spin and orbital polarization of the MBSs. In particular, we focus on the behavior nearby the topological phase transition for the various bands and discuss differences with respect to the canonical spin-Rashba model employed to study topological phase transitions in semiconducting nanowires. Finally conclusions are given in Section V. The Appendix A is devoted to the derivation of the Majorana polarization for the case of multi-orbital topological superconductors, while in Appendix B we report the characterization of the spin-orbital polarization of the states at the Fermi level in the normal phase.




\section{Model and methodology}
\label{sec:model}

In transition metal oxides with perovskite structure the transition metal (TM) elements are surrounded by oxygen (O) in an octahedral environment. Owing to the crystal field potential generated by the oxygen around the TM, the fivefold orbital degeneracy is removed and $d$ orbitals split into the $t_{2 g}$ sector, i.e., $yz$, $zx$, and $xy$, and the $e_g$ sector, i.e., $x^2 - y^2$ and $3z^2 - r^2$.
For a tetragonal symmetry the low-energy electronic structure can be described by a model having only the $t_{2 g}$-orbitals contributing to the Fermi level. Additionally, for weak octahedral distortions, the TM-O bond angle is almost ideal and thus the three $t_{2 g}$-bands are mainly directional and basically decoupled, e.g., an electron in the $d_{xy}$-orbital can predominantly hop along the $y$ or $x$ direction through the intermediate $p_x$- or $p_y$-orbitals. Similarly, the $d_{yz}$- and $d_{zx}$-bands are quasi one-dimensional when considering a square geometry for the 2D TM-O bonding network.  

Furthermore, the atomic spin-orbit interaction (SO) mixes the $t_{2g}$ orbitals thus competing with the quenching of the orbital angular momentum resulting from the crystal field potential. Out-of-plane buckling modes of the TM-O bond are very important in 2DEGs forming at the interface of insulating oxide materials, as they cause an orbital mixing, which is odd in space, of $d_{xy}$ and $d_{yz}$ or $d_{zx}$-orbitals along the $y$ or $x$ directions, respectively.  Indeed, the inversion symmetry breaking is primarily affecting the orbital degrees of freedom and leads to an orbital momemtum locking through the so-called orbital Rashba interaction while the spin-momentum coupling derives from the atomic spin-orbit. The atomic spin-orbit interaction (SO) is then a crucial term to be included into the electronic description both for the setting out of the spin-orbital texture in the reciprocal space \cite{Fukaya2019}, and for the natural mixing of the spin-orbital degrees of the $t_{2g}$-states in competition with the quenching of the orbital angular momentum due to the crystal potential. \\
Since we are interested in the analysis of topological phase that it established as a consequence of time reversal symmetry breaking due to an external magnetic field, the model Hamiltonian we are going to consider includes both a coupling of electron spin and orbital moments to the magnetic field and a superconducting pairing term. The conditions to achieve a topological non-trivial superconducting phase for a quasi one-dimensional nanowire were already discussed in Ref.\cite{Perroni2019}. In particular for the chosen geometry of the nanowire oriented along the $x$-axis, the optimal magnetic field direction for achieving a topological phase is to be perpendicular to the $y$-direction, that is the orientation of the orbital Rashba field.  \\
The model Hamiltonian, including the $t_{2g}$ hopping terms, the atomic spin-orbit coupling, the inversion symmetry breaking term, and the external magnetic field can be generally expressed as \cite{Zhong2013,Khalsa2013,Vivek2017,Fukaya2018,Fukaya2019}
\begin{align}
\mathcal{H}&=\sum_{\bm{k}}\Hat{D}(\bm{k})^{\dagger}H(\bm{k})\Hat{D}(\bm{k}), \\
H(\bm{k})&=H^0+H^\mathrm{SO}+H^{Z}+H^{M} \,,
\end{align}%
where $\Hat{D}^{\dagger}(\bm{k})=\left[ c^{\dagger}_{yz\uparrow \bm{k}}, c^{\dagger}_{zx\uparrow \bm{k}}, c^{\dagger}_{xy\uparrow \bm{k}}, c^{\dagger}_{yz\downarrow \bm{k}}, c^{\dagger}_{zx\downarrow \bm{k}}, c^{\dagger}_{xy\downarrow \bm{k}} \right]$ is a vector whose components are associated with the electron creation operators for a given spin $\sigma$ ($\sigma=[\uparrow,\downarrow]$), orbital $\alpha$ ($\alpha=[xy,yz,zx]$), and momentum $\bm{k}$ in the Brillouin zone. Then $H^0, H^\mathrm{SO}, H^{Z}$ and $H^{M}$ represent the kinetic energy, the spin-orbit, the inversion symmetry breaking and the Zeeman interaction term, respectively. \\
In the spin-orbital basis, $H_0(\bm{k})$ is given by
\begin{align}
&H^0=\Hat{\varepsilon}_{\bm{k}} \otimes \Hat{\sigma}_{0}, \\
&\Hat{\varepsilon}_{\bm{k}}=
\begin{pmatrix}
\varepsilon_{yz} &0 &0 \\
0 & \varepsilon_{zx} &0 \\
0 &0& \varepsilon_{xy}
\end{pmatrix}, \notag \\
&\varepsilon_{yz}=2t_{2}\left(1-\cos{k_x}\right), \notag \\
&\varepsilon_{zx}=2t_{1}\left(1-\cos{k_x}\right), \notag \\
&\varepsilon_{xy}=2t_{1}\cos{k_x}+\Delta_{t}, \notag
\end{align}%
where $\Hat{\sigma}_{0}$ is the unit matrix in spin space and
 $t_{1}$ and $t_{2}$ are the orbital dependent hopping amplitudes.
$\Delta_{t}$ denotes the crystal field potential as due to the symmetry lowering from cubic to tetragonal also related to inequivalent in-plane and out-of-plane transition metal-oxygen bond lengths.
The symmetry reduction yields a level splitting between $d_{xy}$-orbital and $d_{yz}/d_{zx}$-orbitals.
$H^\mathrm{SO}$ denotes the atomic $\bm{l} \cdot \bm{s}$ spin-orbit coupling,
\begin{align}
H^\mathrm{SO}
=\Delta_{\mathrm{SO}}\left[ \Hat{l}_x \otimes \Hat{\sigma}_x+\Hat{l}_y \otimes \Hat{\sigma}_y+\Hat{l}_z \otimes \Hat{\sigma}_z \right],
\end{align}%
with $\Hat{\sigma}_{i}(i=x,y,z)$ being the Pauli matrix in spin space and $\Hat{l}_\alpha$ ($\alpha=x,yz$) are the projections of the $l=2$ angular momentum operator onto the $t_{2g}$ subspace, i.e.,
\begin{align}
\Hat{l}_{x}&=
\begin{pmatrix}
0 & 0 & 0 \\
0 & 0 & i \\
0 & -i & 0
\end{pmatrix}, \\
\Hat{l}_{y}&=
\begin{pmatrix}
0 & 0 & -i \\
0 & 0 & 0 \\
i & 0 & 0
\end{pmatrix}, \\
\Hat{l}_{z}&=
\begin{pmatrix}
0 & i & 0 \\
-i & 0 & 0 \\
0 & 0 & 0
\end{pmatrix},
\end{align}%
assuming $\{d_{yz}, d_{zx}, d_{xy}\}$ as orbital basis.


\begin{figure*}[t]
\includegraphics[height=4.16cm]{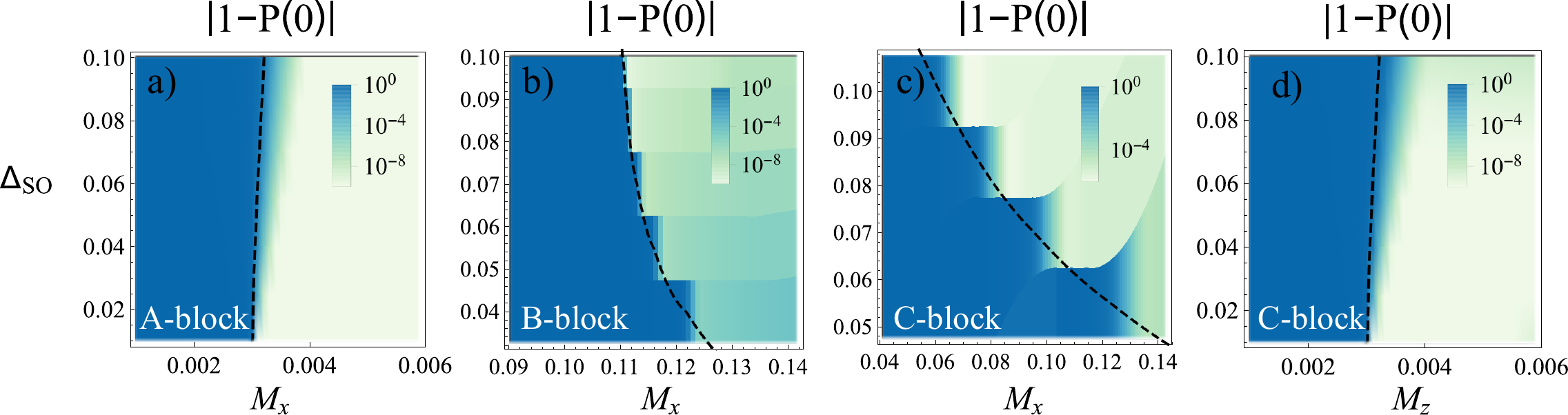}
\caption{\label{fig:phaseDiagram} 
Topological phase diagram evaluated by means of the Majorana polarization $P(\omega=0)$ (see main text for the definition) with a value of about 1 or 0 to signal the onset of a topological or trivial superconducting configuration, respectively. For convenience we indicate as A, B and C the physical cases for a given electron filling that refer to the two bands sector, associated with the $\Gamma$-point Kramers doublet at zero field. The sectors A, B and C occur when moving from lower to higher energies in the spectrum as depicted in Figs. \ref{Fig:1}(a),(b),(c). (a)-(c) topological phase diagram in the spin-orbit coupling/magnetic field plane related to the bands for the A, B, C sectors assuming a magnetic field $M_x$ oriented along the nanowire direction. (d) topological phase diagrama for refers the bands belonging to the block C for an out-of-plane magnetic field $M_z$. The chemical potential has been selected to be pinned at the energy lying in the middle of the split Kramers doublet for each block to distinctively follow the topological behavior of the corresponding orbital sectors. We vary the amplitude of the spin-orbit coupling $\Delta_{SO}$ and the applied magnetic field $M$ to search for the boundary separating the topological and trivial superconducting phase. The black dashed line schematically indicates the transition from a topological to trival superconducting phase as obtained by looking at the gap closing in the momentum space. The gap amplitude for the various orbitals is $\Delta_{\alpha}=0.003$ in unit of $t_1$. All the energies and electronic parameters are in units of $t_1$.}
\end{figure*} 


As mentioned above, the breaking of the mirror plane, due to the out-of-plane offset of the positions of the TM and O atoms, results into an inversion asymmetric orbital Rashba coupling that is described by the term $H^{Z}(\bm{k})$:
\begin{align}
H^{Z}
=\gamma \left( \Hat{l}_y \otimes \Hat{\sigma}_0 \sin{k_x}\right).
\end{align}%
This contribution gives an inter-orbital process, due to the broken inversion symmetry, that mixes $d_{xy}$ and $d_{yz}$ or $d_{zx}$.
Finally, we consider the effects of a magnetic field lying into the plane of the 2D electron gas. The resulting Zeeman-type interaction is described by the Hamiltonian $H^{M}$, which characterizes the coupling of the electron spin and
orbital moments to the magnetic field \cite{Ruhman2014}:
\begin{align}
H^\mathrm{M}
=&M_x \left[  \Hat{l}_x \otimes \Hat{\sigma}_0 + \Hat{l}_0 \otimes \Hat{\sigma}_x \right]+   M_y \left[ \Hat{l}_y \otimes \Hat{\sigma}_0   + \Hat{l}_0 \otimes \Hat{\sigma}_y \right]+\\
&  M_z \left[ \Hat{l}_z \otimes \Hat{\sigma}_0 + \Hat{l}_0 \otimes \Hat{\sigma}_z \right],
\end{align}%
\noindent with $\Hat{l}_0$ being the unit matrix in the orbital space. We notice that the inclusion of the orbital coupling to the field can be neglected because the atomic spin-orbit coupling, once the spin symmetry is broken by the Zeeman field, also acts to orbital polarize the electronic configuration along the same direction \cite{Perroni2019}. 

Concerning the superconducting pairing, we assume that the interaction is local, with spin-singlet symmetry and active only for electrons sharing the same orbital symmetry \cite{Perroni2019,Michaeli2012,Nakamura2013,Loder2013,Mohanta2015}.
Hence, the superconducting term $H^P$ can be expressed as
\begin{equation}
H^P = - g \sum_{{\bf{i}},\alpha} n_{{\bf{i}}\alpha,\uparrow} n_{{\bf{i}}\alpha,\downarrow},
\label{hamilpair}
\end{equation}
where $g$ is the pairing interaction, $n_{{\bf{i}}\alpha,\sigma} =c_{{\bf{i}},\alpha,\sigma}^{\dagger} c_{{\bf{i}}\alpha,\sigma}$ is the local spin-density operator for the $\sigma$ polarization ,and the $\alpha$ orbital, at a given position ${\bf{i}}$ in the square lattice.
We then employ the usual decoupling scheme for the pairing term using a mean-field approach for the spatial and orbital degrees of freedom:
\begin{eqnarray}
H^P & =&  - \sum_{{\bf{i}},\alpha} \Delta_{{\bf{i}},\alpha} \left[ c_{{\bf{i}},\alpha,\uparrow}^{\dagger} c_{{\bf{i}},\alpha,\downarrow}^{\dagger} + c_{{\bf{i}},\alpha,\downarrow}  c_{{\bf{i}},\alpha,\uparrow}  \right]  \nonumber + g \sum_{{\bf{i}},\alpha} D^2_{{\bf{i}},\alpha},
\label{hamilbogo}
\end{eqnarray}
with the pairing amplitude $D_{{\bf{i}},\alpha}=\langle c_{{\bf{i}},\alpha,\downarrow}  c_{{\bf{i}},\alpha,\uparrow}\rangle$ and the order parameter $ \Delta_{{\bf{i}},\alpha}=g D_{{\bf{i}},\alpha}$ are taken in a gauge such as to have a real amplitude. 

For the determination of the topological phase diagram and the spin-orbital character of the Majorana bound states we compute the spectrum and the corresponding eigenvectors of the Bogoliubov- De Gennes Hamiltonian both in the momentum and in real space by exploring different electronic regimes concerning the orbital filling and the spin-orbit strength. 
The numerical tight binding hamiltonian is implemented by using KWANT \cite{Kwant} and solved with the help of NumPy routines \cite{NumPy}. In the following sections we set the parameters of the Hamiltonian in units of the main hopping term $t_1$, specifically as: the weaker hopping amplitude $t_2 = 0.1$, the 
orbital Rashba interaction $\gamma=0.2$,  the tetragonal crystal field potential $\Delta_t=-0.5$, the superconducting pairing $\Delta_{i,\alpha}=0.003$, independent of $i$ and $\alpha$, and the atomic spin-orbit coupling $\Delta_{SO}$ varying from $0.01$ to $0.1$. 
This set of parameters is representative of a physical regime with an electronic hierarchy of the energy scales such that $\Delta_t>\gamma>\Delta_{\mathrm{SO}}$.

\section{Topological phase diagram}
\label{sec:majorana-polarization}

In this section we present the topological phase diagram as a function of the applied Zeeman magnetic field and the strength of the spin-orbit coupling for three representative electron fillings corresponding to the chemical potential crossing the bands nearby the $\Gamma$ point.
For clarity and convenience we indicate as A, B and C each sector of two bands, associated with the $\Gamma$-point Kramers doublet at zero field, which occur when moving from lower to higher energies in the
spectrum (Fig. \ref{Fig:1}(a),(b),(c)). In particular, according to the selected range of parameters for the spin-orbit coupling and crystal field potential, the block A refers to the bands with a dominant $d_{xy}$ character and sub-dominant $(d_{xz},d_{yz})$ contributions (Fig. \ref{Fig:1}(a)). The block B at intermediate energies corresponds to bands arising from Kramers doublets with concomitant highest values of the spin and orbital components (Fig. \ref{Fig:1}(b)). Finally, the high energy bands set out the block C composed of states with dominant $(d_{xz},d_{yz})$ character and a sub-dominant $d_{xy}$ contribution (Fig. \ref{Fig:1}(c)).  
The main purpose is to compare the topological phase diagram for the various bands, with the aim to assess the role of the spin-orbital anisotropy and of the degree of spin-orbital entanglement. 

There are various approaches to identify a topological phase transition where Majorana bound states then occur at the edges of the quantum chain.\cite{CarloRev,SatoAndo,Citro2018} 
As discussed in Ref.~[\onlinecite{Sticlet2012}], the Majorana polarization is one of them being a suitable indicator for detecting the topological phases and it can be considered as a sort of order parameter. 
In analogy with the case of a single band electronic model, one can define the Majorana polarization for a multiband system as follows 
\begin{equation}
P^{L(R)}(\omega)= \sum_n \abs{P_n^{L(R)} (\omega)} \qty[\delta(\omega-e_n)+\delta(\omega+e_n)].
\end{equation}
with
\begin{equation}
P_n^{L(R)}(\omega)= 2 \abs { \sum_{j=1 (N/2+1)}^{N/2 (N)} \sum_{\alpha,\sigma}
u_{n,j,\alpha,\sigma} v_{n,j,\alpha,\sigma} }.
\end{equation} 
Here $N$ is the size of the chain, $e_n$ is a given eigen-energy of the BdG Hamiltonian and $u_{n,j,\alpha,\sigma}$ and $v_{n,j,\alpha,\sigma}$ are respectively the electronic and hole components of the eigenfunctions of the hamiltonian (more details of the derivation are reported in the Appendix A). 

\begin{figure}[t]
\includegraphics[width=0.48\textwidth]{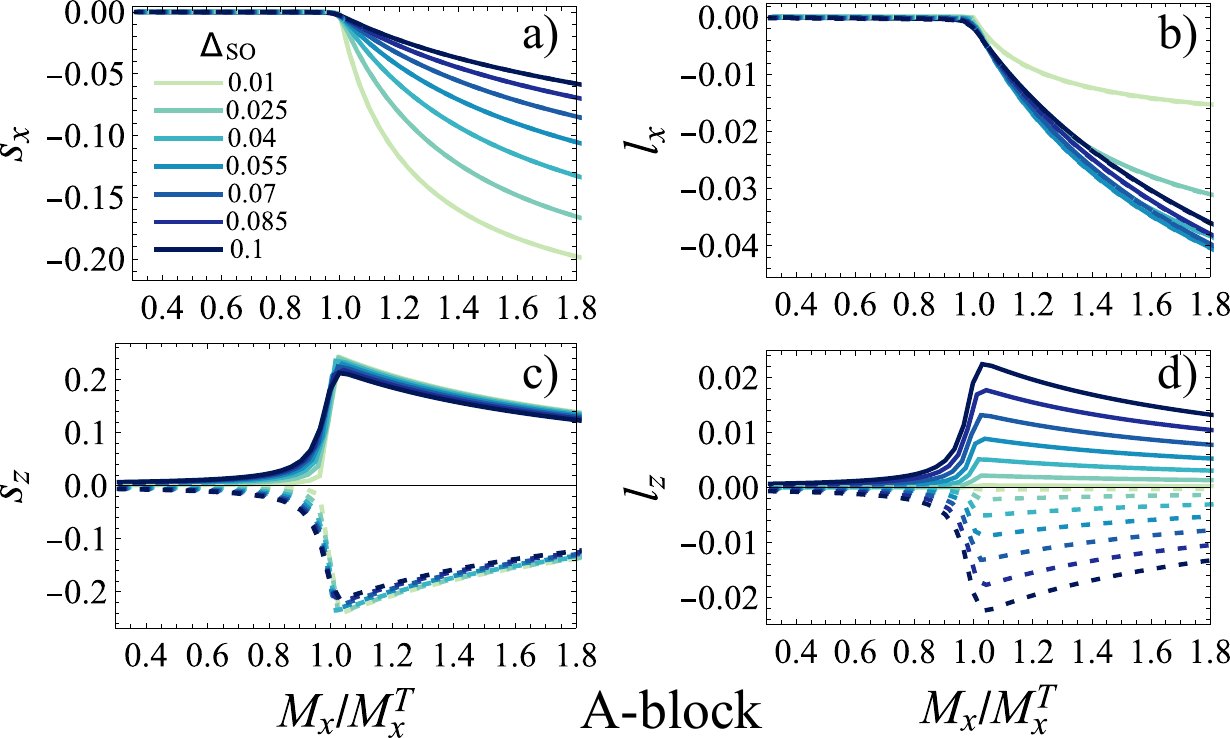}
\caption{$x$- and $z-$ electron components of the spin $s$ ((a) and (c)) and orbital $l$ ((b) and (d)) angular momentum evaluated nearby the topological phase transition point ($M_{x}^{T}$) on the lowest energy excited state in the trivial superconducting phase ($M_x<M_{x}^{T}$) and for the MBS in the topological configuration ($M_x>M_{x}^{T}$). The behavior refers to the topological phase diagram for the electronic states belonging to the lowest energy sector (block A) in the presence of a magnetic field oriented along the nanowire. Solid and dashed lines refers to the MBS spin-orbital component at the two edges of the nanowire. The component collinear to the magnetic field owes the same sign and amplitude at the two edges of the nanowire. The transverse spin and orbital components with respect to the applied field have opposite sign at the two edges but equal amplitude.   
\label{fig:so-pol-bandA}}
\end{figure}

The Majorana polarization has been then used to single out the topological superconducting phase in response to a change of the spin-orbit coupling, by considering different representative electron fillings and an applied Zeeman field along the directions perpendicular to the orbital Rashba field. These field orientations are those more favorable to yield a topological nontrivial state.
Hence, in Fig. \ref{fig:phaseDiagram} we show the topological phase diagram for electron densities corresponding to a chemical potential that uniquely crosses the energy bands in the blocks A, B and C (Fig. 1). 
A common aspect for the phase diagrams linked to the blocks A and C is that the trivial-topological boundary is weakly dependent on the amplitude of the spin-orbit coupling if the field is applied along the easy magnetic axis ($x$ and $z$, respectively), with the critical threshold for the applied magnetic field of the order of the superconducting gap. On the other hand, for the block B and C one needs to apply a magnetic field which is significantly larger than the superconducting gap to induce a topological phase if the magnetic field is applied along the hard magnetic axis. Additionally, the boundary line for the hard magnetic direction (i.e. $x$ for the blocks B and C), as expected, is more sensitive to the strength of the spin-orbit coupling.
Although the phase diagram of the A and C sectors for a field applied along the easy magnetic axis is substantially unchanged by a variation of the spin-orbit coupling, the character of the MBS in the topological phase is strongly dependent on the strength of spin-orbital interaction as we will discuss below in the Sect. IV. 

We notice that the bands A, as pointed out in Ref. \cite{Perroni2019}, show a spin Rashba-like transition with the in-plane magnetic field along the direction of the nanowire ($M_x$) and a transition point approximately determined by $M_x^T \approx \sqrt{\Delta^2+\epsilon_0^2}$, where $\epsilon_0$ is the energy difference between the chemical potential and the bottom of the band. Surprisingly, a similar behaviour with an out-of-plane magnetic field is also observed for the band C although the inversion symmetry splitting deviates from the canonical spin Rashba profile. Also in this case we find that the transition point is essentially determined by $M_z^T \approx \sqrt{\Delta^2+\epsilon_0^2}$.

The validity of using the Majorana polarization to signal the trivial-to-topological phase transition is confirmed by the correspondence of the critical values with those obtained by evaluating the position of the gap closing in the reciprocal space (black dashed lines in Fig. \ref{fig:phaseDiagram}).

\section{Spin-orbital polarization of Majorana bound states}

\begin{figure}[t]
\includegraphics[width=0.48\textwidth]{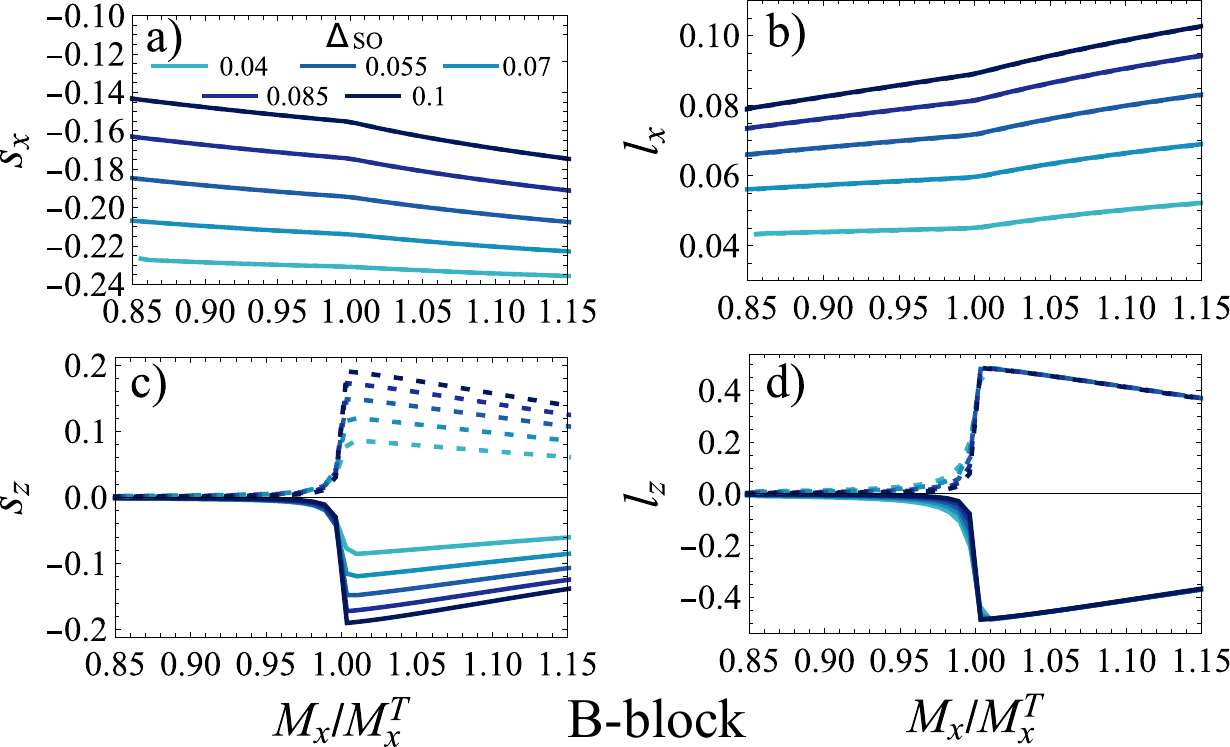} 
\caption{$x$- and $z-$ electron components of the spin $s$ ((a) and (c)) and orbital $l$ ((b) and (d)) angular momentum evaluated across the topological phase transition point ($M_{x}^{T}$) for the lowest energy excited state in the trivial superconducting phase (i.e. for $M_x<M_{x}^{T}$) and for the MBS in the topological side (i.e. for $M_x>M_{x}^{T}$). The behavior refers to the topological phase diagram due to a magnetic field oriented along the nanowire and considering the electronic states of the intermediate energy sector for the normal state spectrum (i.e. bands of the B-block). Solid and dashed lines refers to the MBS spin-orbital component at the two edges of the nanowire. The amplitude is the same for the two MBS localized at the edges while concerning the relative orientation, the component collinear to the magnetic field are parallel while those transverse are anti-aligned.
\label{fig:so-pol-bandB}}
\end{figure} 

\begin{figure}[t]
\includegraphics[width=0.48\textwidth]{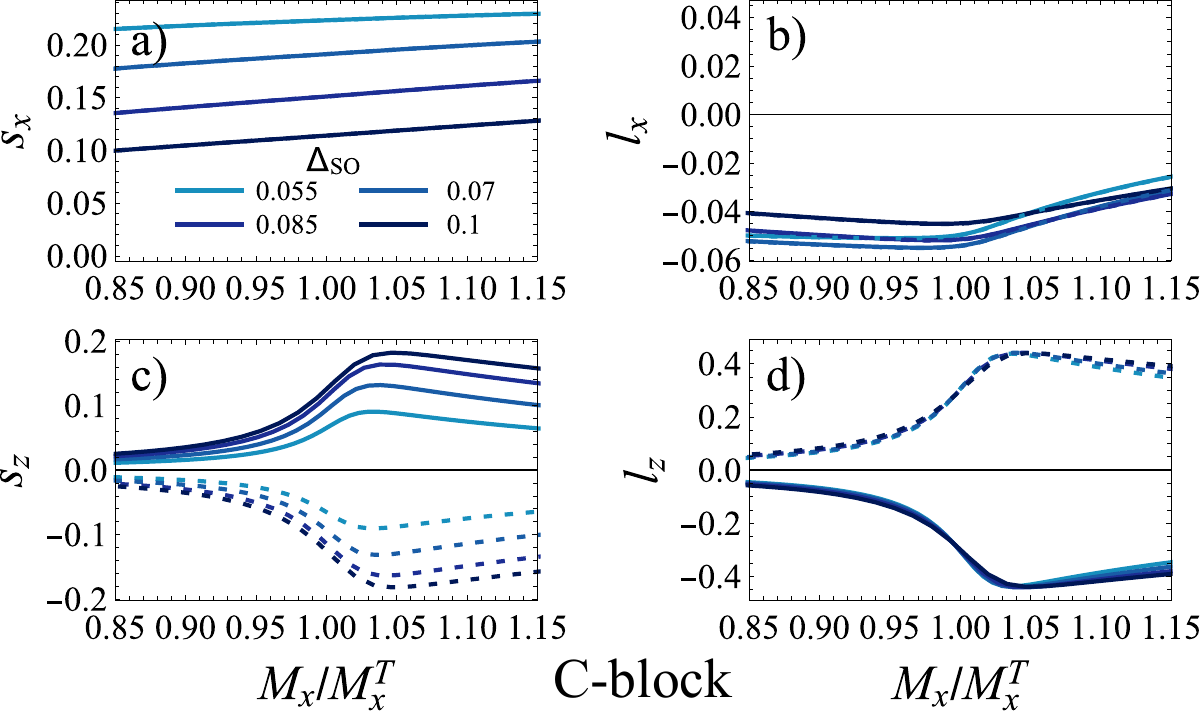} 
\caption{$x$- and $z-$ electron components of the spin $s$ ((a) and (c)) and orbital $l$ ((b) and (d)) angular momentum evaluated across the topological phase transition point ($M_{x}^{T}$) for the lowest energy excited state in the trivial region ($M_x<M_{x}^{T}$) and for the MBS in the topological side ($M_x>M_{x}^{T}$). The behavior refers to the topological phase diagram with an applied magnetic field along the nanowire and for the electronic states belonging to the intermediate energy sector (i.e. block C). Solid and dashed lines refers to the MBS spin-orbital component at the two edges of the nanowire.  
\label{fig:so-pol-bandC}}
\end{figure} 

In this section we consider the behavior of the electron spin-orbital polarization of the MBSs by focusing on the orientation, the spatial profile and the changeover across the phase transition going from in-gap fermionic states to Majorana edge modes. Following the schematic structure of the energy bands, we determine the spin-orbital polarization of the MBSs arising from each band by varying the strength of the spin-orbit coupling. 
There are various questions we aim to address. The first issue to account for is about the dependence of the spin-orbital polarization of MBSs on the strength of the spin-orbital coupling, the orientation of the magnetic field, the spin-orbital anisotropy and on the character of the orbitals which are involved in the pairing at the Fermi level. Additionally, we aim to provide distinctive fingerprints regarding the spin-orbital orientation and the spatial texture of the MBSs at the edge of the superconductor. Another relevant aspect in the problem upon examination is to assess whether the spin-orbital polarization of the electronic states at the Fermi level in the normal state is imprinting the behavior of the spin-orbital content of the MBS.
Due to the intricate spin-orbital character of the electronic states, we expect that the components of the spin-orbital polarization of the MBS are strongly susceptible to the variation of the spin-orbit coupling or the crystal field amplitude in a way that can be markedly orbital dependent. Along this line, the outcome of the analysis unveils striking behaviors of the components of the spin-polarization that are collinear or transverse to the applied magnetic field.

\begin{figure}[t]
\includegraphics[width=0.48\textwidth]{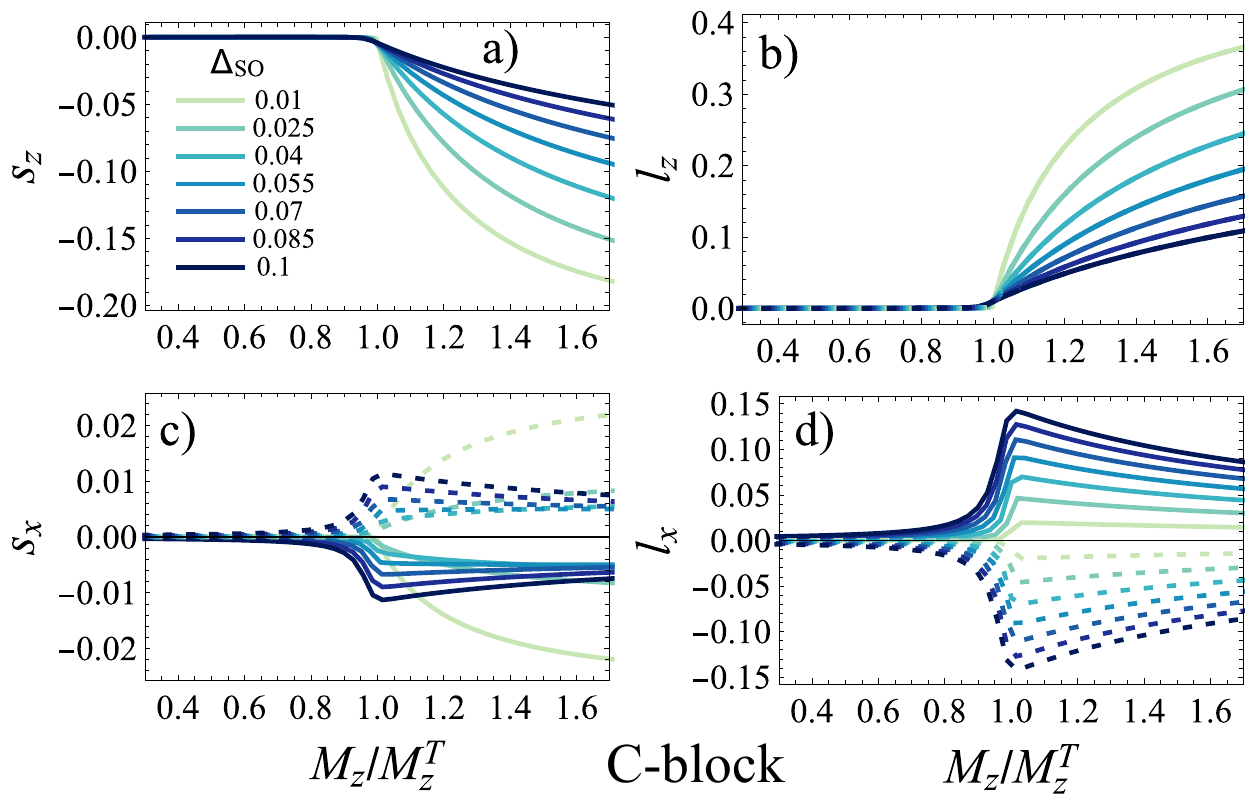} 
\caption{$x$- and $z-$ electron components of the spin $s$ ((a) and (c)) and orbital $l$ ((b) and (d)) angular momentum evaluated nearby the topological phase transition point ($M_{z}^{T}$) for the lowest energy excited state in the trivial region ($M_z<M_{z}^{T}$) and for the MBS in the topological configuration ($M_z>M_{z}^{T}$). The behavior refers to the topological phase diagram with an out-of-plane field oriented along $z$-direction and for the electronic states belonging to the intermediate energy sector (i.e. block C). Solid and dashed lines refers to the MBS spin-orbital component at the two edges of the nanowire.  
\label{fig:so-pol-bandCz}}
\end{figure} 

\begin{figure*}[t]
\includegraphics[width=0.9\linewidth]{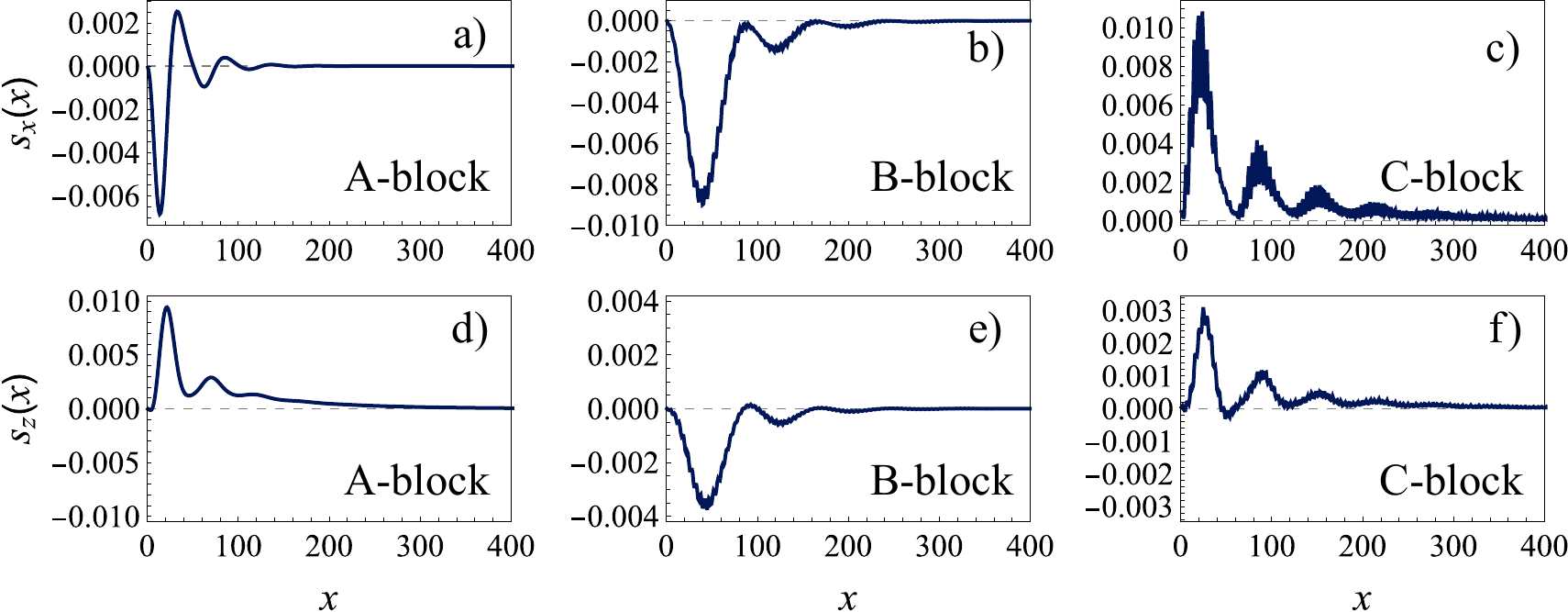} 
\caption{\label{fig:spin_real} 
Spatial profile of the $x$- and $z$- component of the spin density for the MBSs arising from the pairing of electronic states belonging to the sector A ((a) and (d)), B ((b) and (e)), C ((c) and (f)) at a given amplitude of the Zeeman magnetic field along the $x$-direction corresponding to the topological side of the phase diagram. The computation has been performed for a quantum chain with 4000 sites.}
\end{figure*} 

Let us start by considering the topological phase due to an applied magnetic field along the $x$-direction for an electron pairing in the band belonging to the block A with dominant $xy$ orbital character (Fig. \ref{fig:so-pol-bandA}). We observe that the spin-orbital polarization is nonvanishing only for the $x$ and $z$ spatially averaged components. However, the response to a variation of the spin-orbit coupling reveals a remarkable behavior when considering the collinear and transverse to the magnetic field components for the MBS. Indeed, while the $s_x$ spin density gets reduced in amplitude when the spin-orbit coupling increases, the $s_z$ value is essentially unchanged for any spin-orbit value. Furthermore, the $s_z$ spin density has opposite sign components on the two sides of the nanowire while the $s_x$ component has the same sign and value. The fact that the $s_z$ component has a constant amplitude means that any small variation in the spin-orbit coupling leads to a change in the orientation of the electron spin moment of the MBS at each edge of the nanowire.

When considering the electron orbital component of the MBS, the resulting outcome is completely uncorrelated to that of the spin density. The $l_x$ component turns out to be less variable with respect to a change in the spin-orbit coupling unveiling a subtle non monotonous behavior. On the contrary, the $l_z$ projection of the orbital angular moment grows in amplitude with the increase of the spin-orbit strength. As for the spin component, the orbital part has an opposite sign at the two edges for the $z$-orientation while it is collinear for the $x$-direction along the applied magnetic field. 
It is worth pointing out that, although the spin and orbital polarizations are substantially locked at the Fermi level in the normal state configuration through the combination of the orbital Rashba coupling and the spin-orbit interaction (see Appendix B for details), the character of the MBSs unveils a completely opposite behavior. The spin and orbital orientations are essentially misaligned and the misalignment is nontrivially tuned by a change in the strength of the spin-orbital coupling. Another relevant observation of our analysis is that, although the electronic states for the sector A and the resulting topological behavior might be well described by an effective single band spin-Rashba model, the spin-orbital content of the MBS unveils an intricate interplay of the quantum spin and orbital constituents. 

Moving to the topological phase for the states in the B sector, we start by observing that the pairing involves more than one Fermi point and that, due to the magnetic anisotropy, the amplitude of the applied field along the $x$-axis to reach the topological transition has to be larger than that employed for the configurations in the A sector. These elements completely alter the behavior of the $s_x$ and $l_x$ components resulting into a smooth changeover across the topological transition (Fig. \ref{fig:so-pol-bandB}). For the transverse projection ($z$) one has an opposite behavior if compared to the MBSs arising from the bands in the sector A. Indeed, the $l_z$ component is essentially unaffected by the spin-orbit coupling while the $s_z$ density exhibits an increase in the amplitude. This is a remarkable reconstruction of the MBS spin content with an unexpected enhancement of the spin density for a stronger value of the spin-orbit coupling. 
A similar behavior is also obtained when considering the MBSs arising from the topological configuration in the high-energy C block. Again, the orbital component transverse to the applied field is independent of the spin-orbit coupling, while the spin density has a monotonous profile.

We point out that, for the bands of the sector C having an easy orbital axis along the transverse $z$-direction, the spin and orbital contents of the MBSs in Fig. \ref{fig:so-pol-bandCz} are substantially analogous to the ones of the A sector, with the transverse spin component being slightly dependent on the spin-orbit coupling. We attribute this behavior to the different character of the quantum configurations involved in the block C as compared to the A-sector. Indeed, in the block C the two quasi-degenerate orbitals ($xz$,$yz$) with larger amplitude in the electronic configuration are spin-orbit active for the $l_z$ angular momentum component. On the other hand, for the block A the main component of the electronic state is associated only to the $xy$ orbital.  
Another remark is that for the bands of the sector B and C having an easy orbital axis along the transverse $z$-direction, the electron orbital content of the MBSs is substantially constant and robust to changes of the spin-orbital coupling strength.

Finally, we discuss the spatial pattern of the electron spin density of the MBS at the edges of the nanowire (Fig. \ref{fig:spin_real}).  
As suggested by the amplitude of the Majorana polarization, the MBSs wavefunctions are localized at the two edges of the nanowire with a characteristic decaying length of the order of few hundred inter-atomic distances. As expected, the spin and orbital polarizations of the MBSs are non-vanishing only close to the edge where the MBS has a maximal amplitude. We find that the strength of the spin component is comparable for all the band sectors but it can exhibit a sign change as a function of the position thus leading to a non-trivial evolution of the spin orientation in the $xz$ plane. For instance, for the A-block (Fig. 7(a),(d)) the $s_x$ component changes sign moving away from the edge while the $s_z$ component is always positive. Hence, although the spin orientation is pinned in the $xz$ plane, the spatial pattern exhibits a texture with a significant rotation of the spin polarization. This behavior is observed only for the MBS in the A-sector while for the B and C-type MBSs we have a less variable modification of the spin-orientation as a function of the position in the nanowire. It is also worth noticing that some oscillations appear in the real space profile of nonzero components of ${\bm s}$ and ${\bm l}$. These oscillations reflect the characteristic length scales of the Fermi wave-vectors of the normal state configurations. Indeed, the MBSs arising from the A sector show only one harmonic, while for B and C block the MBSs display multiple components for the amplitude modulation in the spatial profile of $s_{x,z}$ and $l_{x,z}$, that can be directly associated to the multiple Fermi points occurring at the corresponding electron filling. This is also confirmed by the analysis of the spectrum in the normal state as discussed in the Appendix B.

\section{Discussion and Conclusions}

We have unveiled the spin-orbital character of MBSs occurring in magnetic-field driven topological superconductors where the electron spin and angular momentum are strongly entangled due to the interplay of atomic spin-orbit interaction and inversion asymmetric couplings (e.g. spin and orbital Rashba).
Taking the paradigmatic example of an electronic system in one-dimension with anisotropic $d$-orbitals (i.e. $d_{xy}$,$d_{zx}$,$d_{yz}$), which is of direct impact for multi-orbitals superconductivity at oxides interface or surface, we find that the resulting MBSs display a rich variety of striking spin-orbital hallmarks. 
A general finding refers to the orientation of the MBS spin-orbital polarization. It is planar and lies in the plane set by the direction of the applied magnetic field and the direction that is transverse to the magnetic and orbital-Rashba fields. While the orientation's plane is common for the spin and angular momentum of the MBS, the spin and orbital polarizations are typically misaligned with an angle that is sensitive to a variation of the electronic interactions. This behavior is in stark contrast with that of the spin-orbital configurations in the normal state, for instance at a given momentum in the Brillouin zone, where the spin and orbital components are essentially collinear. Such observation can be potentially relevant for distinguishing the occurrence of MBSs from conventional in-gap bound states at the edge of the superconductor as induced by inhomogeneities or extrinsic effects. 

The presented analysis allowed us to understand the fundamental interrelation among the spin-orbital polarization of the MBS, the strength of the coupling between the spin and orbital degrees of freedom and the magnetic and orbital anisotropy of the electronic states that contribute to the formation of the topological pairing. For this aim, we have essentially employed the atomic spin-orbit interaction as a knob to modify the degree of the spin-orbital entanglement in order to assess the consequences on the MBS. In this respect, the analysis provides direct access to the spin-orbital susceptibility of the MBS with respect to a variation of the electronic parameters. 
We find that when the magnetic field is applied along an easy axis for the corresponding electronic states at the Fermi level, the transverse to the magnetic field MBS spin-polarization is more resilient to variation of the spin-orbit coupling. We qualitatively attribute this behavior to the fact that the induced transverse component is uniquely tied to the structure of the MBS, since there are no corresponding contributions in the normal state, and that being a hard axis it is weakly activated by a change in the spin-orbit coupling. A completely different behavior is achieved when the topological phase is obtained by applying a magnetic field along a hard axis direction for the paired electrons. In that case, we find that the transverse orbital component to the magnetic field gets substantially unaffected by a modification of the spin-orbit coupling. We argue that this strongly asymmetric behavior can be attributed to the inequivalent orbital susceptibility of the electronic states at the Fermi level. Indeed, the states with dominant $xy$ orbital character the easy axis is in plane, while for those with mixed $xz$,$yz$ orbital configurations the easy axis is along the out-of-plane direction. However, the energy separation of the orbitals due to the crystal field potential makes the out-of-plane direction easier to activate than the in-plane one. Hence, there is a clear separation in the behavior of the spin and orbital degrees of freedom of the MBS and this outcome is expected to occur for spin-orbital correlated superconductors in a regime where the spin-orbit interaction competes with the structural couplings.

Although the analysis focused on the role of the spin-orbit interaction we argue that similar response can also occur when other electronic parameters are varied especially if considering the crystal field potential associated to the structural distortions. Indeed, the tetragonal splitting of the $d$-orbitals typically tends to quench the orbital angular momentum and thus competes with the spin-orbit coupling by indirectly reducing its effectiveness. 
In this context, we point out that a modification of the structural configuration through local strains or by applying electric field would manifest into striking effects in terms of reorientation of the spin-orbital polarization of the MBS. Remarkably, the rearrangement of the spin-orbital MBS is different for the spin and orbital components and it also manifests with a complete restructuring of the spatially resolved textures.

From an experimental point of view, we argue that the identified signatures of the MBSs can be accessed in spin-selective transport probes and would manifest in a strong anisotropic response of the conductance. Since the averaged spin-polarization of the MBSs can be sensitive to structural changes, we also expect a significant strain driven magnetic anisotropy to occur in the zero bias tunneling conductance. A similar response would be detected in spin resolved STM experiments where the atomic profile of the MBS spin polarization can be directly accessed. Our prediction of spatial dependent orientation of the spin-polarization of the MBS with a gradient that is sensitive to the orbital character of the paired electrons can be employed to assess the nature of the topological phases in multi-orbital superconductors.
Finally, concerning the orbital polarization of the MBS, we point out that it is much challenging to design an experiment to directly access the orbital angular momentum of the MBS. While the anisotropy of the $d$-orbitals would naturally manifest into a characteristic angular dependence of the tunneling or STM conductance, in order to pin point the orbital polarization one would device a specific filter of orbital-selective angular momentum. One possible way out is to design a tunnel barrier with tunable inversion asymmetric interactions (e.g. Rashba and Dresselhaus) that, due to the orbital momentum locking, can allow the injected control of electrons with selected orbital polarization along a given direction. This setup definitely requires a high degree of control of interface and materials engineering. Advancements along this direction might open the path to a fully spin-orbital spectroscopic probe of in-gap modes of topological superconductors and contribute to single-out distinctive signature for the experimental detection of MBS with strong interplay of spin-orbital degrees of freedom.


\begin{acknowledgements}
We acknowledge discussions with Anton Akhmerov, Marco Salluzzo, Francesco Romeo and Alfonso Maiellaro.
This work was supported by the project Quantox of QuantERA-NET Cofund in Quantum Technologies, implemented within the EU-H2020 Programme and the project ``Two-dimensional Oxides Platform for SPIN-orbitronics nanotechnology (TOPSPIN)" funded by the MIUR-PRIN Bando 2017 - grant 20177SL7HC.
\end{acknowledgements}

\appendix

\section{Majorana polarization}
The real space creation and annihilation operators can be expressed in the basis of single particle eigenfunctions of the BdG Hamiltonian in the following form:  
\begin{equation}
\label{eq:fieldOp1}
 \hat\Psi_{j,\alpha,\sigma}(t)=\sum_{n} \left( u_{n,j,\alpha,\sigma} \hat {c}_{n,\alpha,\sigma}(t)+v_{n,j,\alpha,\sigma} \hat {c}^\dagger_{n,\alpha,\sigma}(t) \right) 
\end{equation}
\begin{equation}\label{eq:fieldOp2}
\hat{\Psi}^\dagger_{j,\alpha,\sigma}(t)=\sum_{n} \left( u^*_{n,j,\alpha,\sigma} \hat {c^\dagger}_{n,\alpha,\sigma}(t)+v^*_{n,j,\alpha,\sigma} \hat {c}_{n,\alpha,\sigma}(t) \right),
\end{equation}
where $ u_{n,j,\alpha,\sigma}$ and $v_{n,j,\alpha,\sigma}$ are, respectively, the electronic and hole component of the $n$-th eigenfunction in the orbital $\alpha$ and spin $\sigma$, calculated at the position $j$. 

The most generic Majorana operators can be written as 

\begin{eqnarray}\label{eq:majOp1}
 \hat {\gamma}^a_{j,\alpha,\sigma}(t)&&=e^{\imath \varphi} \hat\Psi_{j,\alpha,\sigma}(t)+
 e^{-\imath \varphi} \hat{\Psi}^\dagger_{j,\alpha,\sigma}(t)
\end{eqnarray} 

\begin{eqnarray}\label{eq:majOp2}
 \hat {\gamma}^b_{j,\alpha,\sigma}(t)&&=\imath ( e^{\imath \varphi} \hat\Psi_{j,\alpha,\sigma}(t)-e^{-\imath \varphi} \hat{\Psi}^\dagger_{j,\alpha,\sigma}(t) ).
\end{eqnarray}

Majorana polarization has been introduced in \cite{Stoudenmire2011,Chevallier2012} in order to detect the topological phases \cite{Sticlet2012,Perfetto2013,Sedlmayr2015a,Sedlmayr2015,Bena2017a}. It can be interpreted as the difference of the probabilities of having a Majorana modes $\hat {\gamma^a}$ and $\hat {\gamma^b}$, at position $j$, in the orbital $\alpha$, with spin $\sigma$ and energy $\omega$. In the language of Green's functions, it is related to the two local spectral functions:  
\begin{equation}\label{eq:polSF}
    P_{j,\alpha,\sigma}(\omega)=A^{a}_{j,\alpha,\sigma}(\omega)-A^{b}_{j,\alpha,\sigma}(\omega)
\end{equation}
with 
\begin{eqnarray*}\label{eq:specFunc}
A^{a,b}_{j,\alpha,\sigma}(\omega)&& = \\&&
- \frac{1}{\pi} \Im \qty[ -\imath \int_{-\infty}^{\infty} e^{\imath \omega t} \theta(t) 
\expval{\acomm { \hat{\gamma}^{a,b}_{j,\alpha,\sigma}(t) } {\hat{\gamma}^{a,b}_{j,\alpha,\sigma}(0)} }].
\end{eqnarray*}
We can write the anti-commutator of Eq.~\ref{eq:specFunc} using Eq.s ~\ref{eq:majOp1} and~\ref{eq:majOp2}, obtaining:

\begin{eqnarray*}
&&\acomm { \hat{\gamma}^{a}_{j,\alpha,\sigma}(t) } {\hat{\gamma}^{a}_{j,\alpha,\sigma}(0)}= \acomm{\hat{\Psi}_{j,\alpha,\sigma}(t)}{\hat{\Psi}^\dagger_{j,\alpha,\sigma}(0)} + \\&& 
\acomm{\hat{\Psi}^\dagger_{j,\alpha,\sigma}(t)}{\hat{\Psi}_{j,\alpha,\sigma}(0)}
+ e^{2 \imath \varphi} \acomm{\hat{\Psi}_{j,\alpha,\sigma}(t)}{\hat{\Psi}_{j,\alpha,\sigma}(0)}
+ \\&& e^{-2 \imath \varphi} \acomm{\hat{\Psi}^\dagger_{j,\alpha,\sigma}(t)}{\hat{\Psi}^\dagger_{j,\alpha,\sigma}(0)}
\end{eqnarray*}
and 
\begin{eqnarray*}
&&\acomm { \hat{\gamma}^{b}_{j,\alpha,\sigma}(t) } {\hat{\gamma}^{b}_{j,\alpha,\sigma}(0)}= \acomm{\hat{\Psi}_{j,\alpha,\sigma}(t)}{\hat{\Psi}^\dagger_{j,\alpha,\sigma}(0)}+ \\ &&
\acomm{\hat{\Psi}^\dagger_{j,\alpha,\sigma}(t)}{\hat{\Psi}_{j,\alpha,\sigma}(0)}+ \\&&
- e^{2 \imath \varphi} \acomm{\hat{\Psi}_{j,\alpha,\sigma}(t)}{\hat{\Psi}_{j,\alpha,\sigma}(0)}+ \\&&
- e^{-2 \imath \varphi} \acomm{\hat{\Psi}^\dagger_{j,\alpha,\sigma}(t)}{\hat{\Psi}^\dagger_{j,\alpha,\sigma}(0)}.
\end{eqnarray*}

The two {\it non-anomalous} terms are wiped out by the difference in Eq.~\ref{eq:polSF}, which results in
\begin{widetext}
\begin{equation}
P_{j,\alpha,\sigma}(\omega)=-\frac{1}{\pi} \Im 
\qty[ -\imath \int_{-\infty}^{\infty} 2 e^{\imath \omega t} \theta(t) 
\qty( e^{2 \imath \varphi} \expval{\acomm{\hat{\Psi}_{j,\alpha,\sigma}(t)}{\hat{\Psi}_{j,\alpha,\sigma}(0)}}
+e^{-2 \imath \varphi} \expval{\acomm{\hat{\Psi}^\dagger_{j,\alpha,\sigma}(t)}{\hat{\Psi}^\dagger_{j,\alpha,\sigma}(0)}} ) ].
\end{equation}
\end{widetext}

Both anti-commutators can be rewritten in terms of single particle eigenfunction of the hamiltonian:
\begin{equation}
\expval{\acomm{\hat{\Psi}_{j,\alpha,\sigma}(t)}{\hat{\Psi}_{j,\alpha,\sigma}(0)}}= \sum_{n}u_{n,j,\alpha,\sigma}v_{n,j,\alpha,\sigma}\qty(e^{\imath e_n t}+e^{-\imath e_n t} )     
\end{equation}
and  
\begin{equation}
\expval{\acomm{\hat{\Psi}^\dagger_{j,\alpha,\sigma}(t)}{\hat{\Psi}^\dagger_{j,\alpha,\sigma}(0)}}= 
\sum_{n}u^*_{n,j,\alpha,\sigma}v^*_{n,j,\alpha,\sigma}\qty(e^{\imath e_n t}+e^{-\imath e_n t} ).
\end{equation}
By performing the integration, we can write:
\begin{widetext}
\begin{multline}
P_{j,\alpha,\sigma}(\omega)=-\frac{2}{\pi} \Im \left\{ \sum_{n} e^{2 \imath \varphi} u_{n,j,\alpha,\sigma}v_{n,j,\alpha,\sigma} \qty[P\qty(\frac{1}{\omega-e_n})+P\qty(\frac{1}{\omega+e_n})-\imath \pi \qty(\delta(\omega-e_n)+\delta(\omega+e_n)) ] \right. + \\ 
\left.  e^{-2 \imath \varphi} u^*_{n,j,\alpha,\sigma}v^*_{n,j,\alpha,\sigma}\qty[P\qty(\frac{1}{\omega-e_{n}})+P\qty(\frac{1}{\omega+e_{n}})-\imath \pi \qty(\delta(\omega-e_{n})+\delta(\omega+e_{n})) ] \right\}
\end{multline}
\end{widetext}
and therefore

\begin{eqnarray*}
P_{j,\alpha,\sigma}(\omega)=-\frac{2}{\pi} \Im [ \sum_{n,\alpha,\sigma} 
\Re \left[ e^{2 \imath \varphi} u_{n,j,\alpha,\sigma} v_{n,j,\alpha,\sigma} \right] 
\qty[P\qty(\frac{1}{\omega-e_n})+
P\qty(\frac{1}{\omega+e_n})-\imath \pi \qty(\delta(\omega-e_n)+\delta(\omega+e_n)) ] ].
\end{eqnarray*}

Then, by taking the imaginary part, one obtains 
\begin{widetext}
\begin{equation}
P_{j,\alpha,\sigma}(\omega)= 2 \sum_n \Re \left[e^{2 \imath \varphi} u_{n,j,\alpha,\sigma} v_{n,j,\alpha,\sigma} \right]  \qty[\delta(\omega-e_n)+\delta(\omega+e_n)].
\end{equation}
\end{widetext}
The integral of MP in the whole Hilbert space of a closed system is zero \cite{Bena2017a}, but if two Majorana states are spatially separated, the integral in each separated region is equal to $1$. For this reason, in the main text we have calculated the integral of MP in the left and right half of the wire, by summing on spin and orbital degrees of freedom
\begin{widetext}
\begin{equation}
P^{L(R)}(\omega)= \sum_n 2 \Re \left[ \sum_{j=1 (N/2+1)}^{N/2 (N)} \sum_{\alpha,\sigma}
e^{2 \imath \varphi} u_{n,j,\alpha,\sigma} v_{n,j,\alpha,\sigma} \right]
\qty[\delta(\omega-e_n)+\delta(\omega+e_n)] = \sum_n P_n^{L(R)} (\omega) \qty[\delta(\omega-e_n)+\delta(\omega+e_n)].
\end{equation}
\end{widetext}
The quantity $P_n(\omega)$ is the real part of a complex number whose phase depends on the particular choice of the global wavefunction phase factor. Therefore, the only physically relevant quantity, for each eigenstate labelled by $n$, is the modulus of $P_n(\omega)$, resulting in the final definition: 
\begin{equation}
P^{L(R)}(\omega)= \sum_n \abs{P_n^{L(R)} (\omega)} \qty[\delta(\omega-e_n)+\delta(\omega+e_n)].
\end{equation}
with
\begin{equation}
P_n^{L(R)}(\omega)= 2 \abs { \sum_{j=1 (N/2+1)}^{N/2 (N)} \sum_{\alpha,\sigma}
u_{n,j,\alpha,\sigma} v_{n,j,\alpha,\sigma} }.
\end{equation}

\begin{figure*}[t!]
\centering
\includegraphics[height=9.2cm,width=0.7\textwidth]{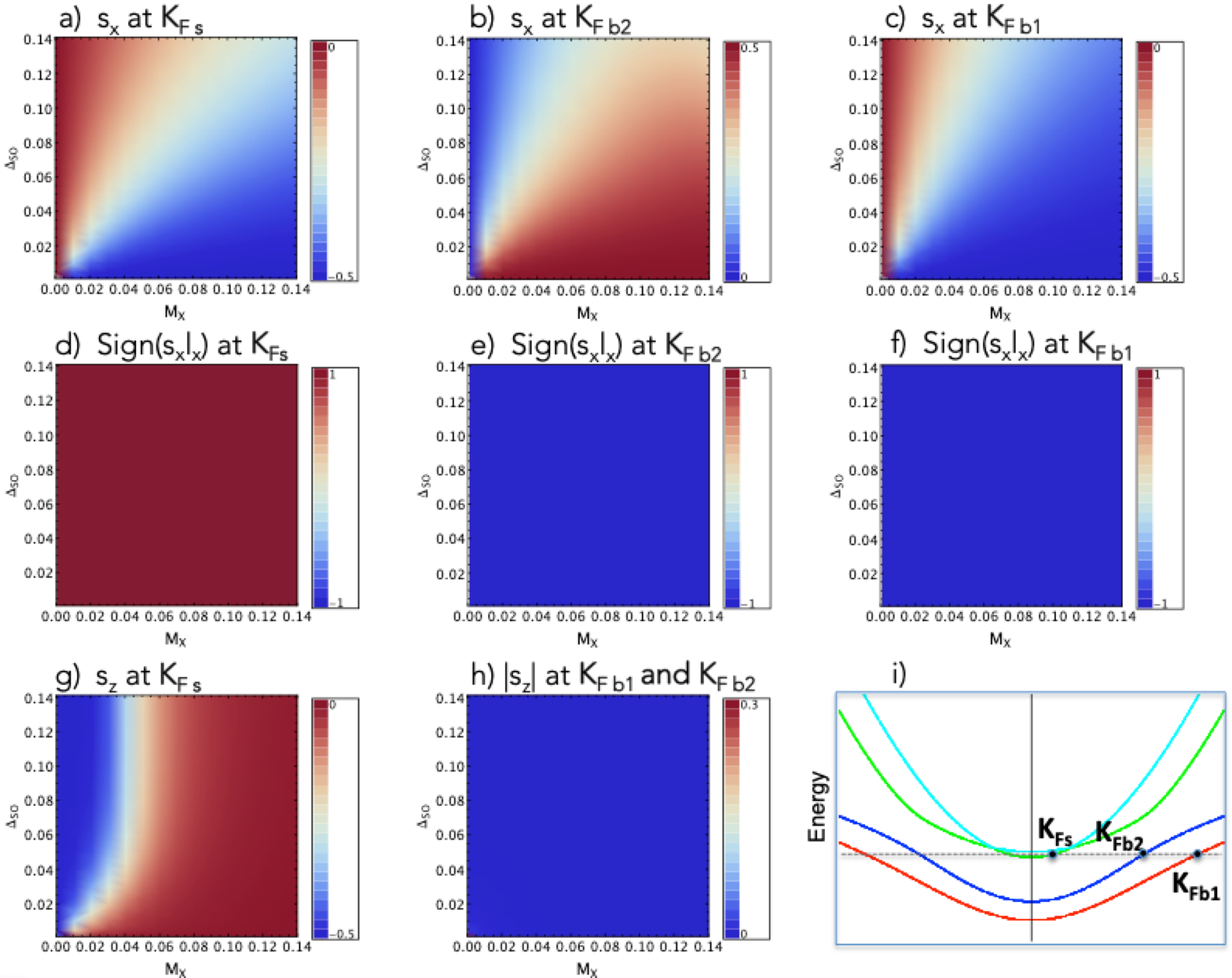}
\protect\caption{Density plots representing the dependence upon the spin-orbit coupling $\Delta_{SO}$ and the Zeeman field $M_x$ of the average spin polarization at the Fermi level in the normal state, for a choice of the chemical potential which is depicted in panel i). In panels d), e) and f) the sign of the product of the components $s_xl_x$ is reported for the three distinct Fermi points.}
\label{NormalPolarizationMap}
\end{figure*}

\section{Spin-orbital polarization at the Fermi level in the normal phase}
In this Appendix we investigate the spin-orbital character of the electronic states at the Fermi level in the normal phase of the model described by the Hamiltonian in Eq. 1. We solve such a model by imposing periodic boundary conditions along the wire direction $x$. The emerging electronic band structure is made up by three blocks with inequivalent orbital character, A, B and C, each forming a Kramers doublet at the $\Gamma$ point, as shown in  Fig. 1(a),(b),(c). Depending on the choice of the electron filling, one or several bands cut the Fermi level, thus leading to the presence of multiple Fermi points $K_F$. Here we will focus on the representative case corresponding to the chemical potential crossing the bands of the B block nearby the $\Gamma$ point. In such a case, the Fermi points occur at characteristic vectors defined as $\pm K_s$, $\pm K_{Fb1}$ and $K_{Fb2}$, which arise from the the lowest of the field split bands of the B sector, and from the highest and lowest bands of of the A sector, respectively. This circumstance is graphically depicted in Fig.\ref{NormalPolarizationMap} i).\\
Figure \ref{NormalPolarizationMap} shows a comprehensive overview of the evolution of the average spin polarization in the $(x,z)$ plane at the different $K_F$ vectors, upon variation of the spin-orbit coupling $\Delta_{SO}$ and of the Zeeman field $M_x$.
We recall that the intricate spin-orbital entanglement of the electronic states yields a strong anisotropy for the magnetic response of the bands under consideration. In particular, in the adopted regime of parameters for the spin-orbit coupling and crystal field potential, the Kramers doublet of the A block is marked by a nonvanishing average spin density both along $x$ and $z$ directions, while the B states are characterized by the highest values of the spin components along $z$. This implies the existence of hard/easy spin directions, specifically $x$ is easy from the bands of block A while it is hard for the bands of block B.\\
The $s_x$ polarization, i.e. collinear to the applied field is shown in panels a), b) and c) in Fig.\ref{NormalPolarizationMap}. From the inspection of the figure, we observe that the spin component along the direction of the field has always a monotonous evolution with $\Delta_{SO}$ and $M_x$. It is evident that for  all the states at each $K_F$, the $s_x$ component grows in absolute value with the Zeeman field and is strongly sensitive to the variation of the spin-orbit strength, reducing in amplitude with increasing $\Delta_{SO}$. Beyond such similar monotonous behavior, we point out some important differences which markedly depend on the specific spin-orbital sector. We notice that the spin polarization is more susceptible to the variation of the spin-orbit coupling for the B state at $K_{Fs}$. Moreover, we observe that the states of the A block are characterized by an opposite sign of the spin polarization, being parallel and antiparallel to the applied field.\\
The $s_z$ component, which is orthogonal to the Zeeman field, is zero by symmetry. In our calculations, we consider a small symmetry breaking field along this direction and observe that the value of $s_z$ is essentially independent on the spin-orbit coupling, as shown in Fig.\ref{NormalPolarizationMap}. g) and h). The spin density arising from the bands of the A block is always vanishing. On the other hand, in the B block we distinguish two regimes: the polarization is maximum in a small window below a threshold of the magnetic field which is almost independent on $\Delta_{SO}$, while its is vanishing above this window.\\
Finally,  in panels d), e) and f) we display the sign of the product ($s_xl_x$), which is representative of the relative orientation of the components of the spin and orbital angular momenta collinear to the magnetic field. It is evident that the sign is uniform in the parameter space, but shows unalike behavior for the distinct states at the Fermi level. At $K_{Fs}$ the spin and angular momentum turn out to be parallel, while they have an opposite sign in the case of the states belonging to the A block. Such behavior reflects the different orbital susceptibiliy of the states belonging to the A and B sectors;  $l_x$ is unquenched and substantial, also being antiparallel to $s_x$ at the $\Gamma$ point, within the Kramers doublet of the A block. For the states of the B block, $l_x$ is identically zero at the $\Gamma$ point, and the effect of the field is to induce a small nonvanishing component, which is parallel to the spin polarization.


\end{document}